\documentclass[aps,showpacs,prb,twocolumn,amsmath,amssymb,superscriptaddress]{revtex4-1}
\usepackage{amssymb}
\usepackage{graphicx}
\usepackage{dcolumn}
\usepackage{bm}

\begin{document}

\bibliographystyle{apsrev}

\newcommand{\bs}{{\bm \sigma}}
\newcommand{\BS}{Bi$_2$Se$_3$}

\newcommand{\la}{\langle}
\newcommand{\ra}{\rangle}
\newcommand{\da}{\dagger}
\newcommand{\D}{d^2{\bf r}}
\newcommand{\norm}{\frac{1}{\sqrt{V}}}
\newcommand{\up}{\uparrow}
\newcommand{\down}{\downarrow}

\newcommand{\lc}{\lowercase}

\newcommand{\no}{\nonumber}
\newcommand{\be}{\begin{equation}} 
\newcommand{\ee}{\end{equation}}
\newcommand{\bea}{\begin{eqnarray}} 
\newcommand{\eea}{\end{eqnarray}}

\newcommand{\vQ}{{\bf Q}}
\newcommand{\vk}{{\bf k}}
\newcommand{\vq}{{\bf q}}
\newcommand{\vM}{{\bf M}}
\newcommand{\vm}{{\bf m}}
\newcommand{\vH}{{\bf H}}
\newcommand{\vh}{{\bf h}}
\newcommand{\vvr}{{\bf r}}
\newcommand{\vu}{{\bf u}}
\newcommand{\hz}{\hat{{\bf z}}}
\newcommand{\hx}{\hat{{\bf x}}}
\newcommand{\hq}{\hat{{\bf q}}}
\newcommand{\hk}{\hat{{\bf k}}}
\newcommand{\tq}{\tilde{{\bf q}}}
\newcommand{\tk}{\tilde{{\bf k}}}
\newcommand{\tmu}{\tilde{\mu}}

\newcommand{\hv}{\hat{{v}}}
\newcommand{\hn}{\hat{{\bf n}}({\bf r})} 
\newcommand{\thk}{\vartheta_{\bf k}}
\newcommand{\hthk}{\frac{\vartheta_{\bf k}}{2}}

\newcommand{\Q}{($0,0,\frac{2\pi}{c}$)}


\title{Surface phonon propagation in topological insulators}

\author{Peter Thalmeier}
\affiliation{Max Planck Institute for Chemical Physics of Solids,
01187 Dresden, Germany}
\date{\today}
\begin{abstract}
The effect of helical Dirac states on surface phonons in a topological insulators is investigated. Their coupling is derived in the continuum limit by assuming displacement dependent Dirac cones. The resulting renormalisation of sound velocity and attenuation and its dependence on chemical potential and wave vector is calculated. At finite wave vectors a Kohn anomaly in the renormalized phonon frequency is caused by intraband-transitions. It appears at wave vectors $q<2k_F$ due to a lack of backscattering for helical Dirac electrons. The wave vector and chemical potential dependence of this anomaly is calculated.

\end{abstract}

\pacs{73.20.-r, 43.35.+d, 63.20.kd}

\maketitle


\section{Introduction}
\label{sect:introduction}

Strong topological insulators \cite{Hasan10} have metallic surface electron states which are protected by time reversal invariance \cite{Fu07}. One of the most prominent examples is Bi$_2$Se$_3$ which has a single Dirac cone with a linear dispersion at the zone center of the surface Brillouin zone. Due to the strong spin orbit coupling the spin is locked perpendicular to the wave vector for each surface state. These features have been confirmed by photoemission \cite{Hsieh09,Chen10a,Wray10}. The chemical potential may be tuned by surface or bulk doping \cite{Chen10a} or application of a gate voltage \cite{Checkelsky10,Chen10b}. Apart from photoemission it has been difficult to find the signature of surface states in transport results which may be masked by bulk contributions \cite{Butch10}. It is therefore desirable to use methods that probe naturally only the surface states. One possible method  is to study the propagation of surface acoustic waves (SAW) by ultrasonic means. This method has been used successfully for the conventional two dimensional electron gas in GaAs heterostructures with parabolic dispersion \cite{Wixforth86,Willett90} and it has also been proposed for studying magnetoconductance of Dirac cone quasiparticles in graphene \cite{Thalmeier10}.\\

Here we investigate the surface acoustic phonon propagation in topological insulators, using the continuum limit for the surface conduction state and the phonon modes. This means a linear dispersion in both cases given by Fermi velocity $v_F$ and sound velocity $v_s$ respectively. Experimentally so far only bulk optical phonons have been investigated in Bi$_2$Se$_3$ \cite{LaForge10,Sushkov10}.
The renormalisation of surface acoustic phonons by helical Dirac electrons should carry a signature of the latter, i.e., their linear density of states around the Dirac point, the linear dependence of Fermi wave number on chemical potential and the suppressed backscattering due the spin locking perpendicular to the wave vector. The latter should play an important role at finite phonon wave vector where the phonon frequency is renormalised by both intraband and interband contributions.\\

In Sect.~\ref{sect:helicalmodel} we give the basic properties of the helical Dirac fermions. In Sect.~\ref{sect:coupling} their coupling to surface phonons is derived and in Sect.~\ref{sect:phonon} the renormalised sound velocity, attenuation and Kohn anomalies in phonon frequency are calculated. Finally Sect.~\ref{sect:conclusion} gives the conclusions.

\section{Helical Dirac fermion model}
\label{sect:helicalmodel}

The protected surface states of a 3D strong topological insulator form a single Dirac cone located at a time reversal invariant point in the Brillouin zone. Their low energy (as compared to the bulk valence band gap) effective Dirac Hamiltonian may be written as \cite{Zhou09,Feng10,Liu10}
\bea
H_0=iv_F\int\D\psi^\da(\vvr)\bs\cdot(\hz\times\nabla)\psi(\vvr)
\label{eq:HD}
\eea
Here $v_F$ is the Fermi velocity, $\bs,\hz$ are the electron spin and the fixed surface normal respectively. The quantisation axis for the electron spin is also chosen parallel to the surface normal. The free massless Dirac electrons are represented by the two component spinors
\be
\psi(\vvr)=\norm\int\D e^{-i\vk\cdot\vvr}\psi_{\vk}; \quad
\psi_{\vk}=
\left(
\begin{array}{c}
c_{\vk\up}\\
c_{\vk\down}
\end{array}
\right)
\ee
Adding a term including the chemical potential $\mu$ the diagonalsation leads to the Dirac cone energies
\be
\epsilon_{\vk\tau}=\pm v_F|\vk|-\mu
\ee
where  $|\vk|=(k_x^2+k_y^2)^\frac{1}{2}$ and $\tau=\pm 1$ is the helical pseudospin index denoting the postive or negative spin helicity of the eigenstates. They are obtained from the free electron states by the unitary transformation  matrix
\begin{equation}
W=\frac{1}{\sqrt{2}}\left(
\begin{array}{cc}
e^{i\hthk} & ie^{-i\hthk} \\
e^{i\hthk}&  -ie^{-i\hthk}   
\end{array}
\right)
\end{equation}

Which leads to the creation and annihilation operators for helical eigenstates $\gamma^\da_{\vk\tau}$ according to
\bea
\gamma^\da_{\vk\pm}&=&\frac{1}{\sqrt{2}}
(e^{-i\hthk}c^\da_{\vk\up}\mp i e^{i\hthk}c^\da_{\vk\down})\no\\
\gamma_{\vk\pm}&=&\frac{1}{\sqrt{2}}
(e^{i\hthk}c_{\vk\up}\pm i e^{-i\hthk}c_{\vk\down})
\label{eq:helical}
\eea
Here  the rotation angle $\thk$ to helical eigenstates is given by $\tan\thk = k_x/k_y$. The existence of surface helical states in the Dirac cone has been proven most clearly for the \BS~ compound by ARPES experiments \cite{Chen10a,Wray10}. As mentioned in the introduction further evidence for their physical effects should be primarily sought with other surface specific probes.

\section{Coupling of surface phonon modes to the Dirac cone}
\label{sect:coupling}

One possibility is the investigation of long wave length surface phonons, in particular surface acoustic waves (SAW). The latter has been very successfully used for the study of the 2D electron gas with parabolic dispersion found in GaAs heterostructures \cite{Wixforth86,Willett90} and it has also been proposed as a method to investigate the chiral Dirac electrons in graphene monolayers \cite{Thalmeier10}, in particular the effect of Landau quantisation. Similar considerations may be fruitful for the surface Dirac states of 3D topological insulators. In the previous cases a phenomenological approach was used where the modification of surface phonon properties is expressed through the conductivity tensor. Here we prefer a more microscopic approach to investigate the effect of the unconventional helical spin polarisation of surface states in the topological insulator.\\ 

%
\begin{figure}
\includegraphics[width=80mm]{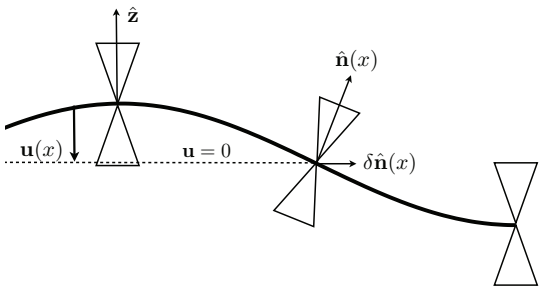}
\caption{Schematic view of adiabatic modification of helical surface states by a surface phonon in the xz- plane. The surface normal
$\hat{{\bf n}}(x)=\hat{{\bf z}}+\delta\hat{{\bf n}}(x)$ giving the local orientation of the Dirac cone is periodically modulated by the surface displacement amplitude \vu(x).}
\label{fig:Fig1}
\end{figure}
%

We first derive a suitable low energy and long wave length effective Hamiltonian for the coupling of surface phonons and helical Dirac states. This will be done by using an adiabatic generalisation of the unperturbed Hamiltonian in Eq.~(\ref{eq:HD}) valid for $v_s\ll v_F$ where $v_s$ is the sound velocity. This  can be achieved by replacing the global surface normal $\hz$ by the local normal $\hn$. Its dependence on surface position $\vvr =(x,y)$  is determined by the amplitude of the surface phonon mode. This leads to the modified Hamiltonian
\bea
H=iv_F\int\D\psi^\da(\vvr)\bs\cdot(\hn\times\nabla)\psi(\vvr)
\label{eq:HL}
\eea
For small phonon wave vectors ($q\ll k_0=\pi/a$, a = lattice constant) and amplitudes we may expand $\hn=\hz+\delta\hn$ where $\delta\hn$ is the local deviation from the global surface normal due to the phonon mode. The geometry of this deformation is schematically shown in Fig.~\ref{fig:Fig1}. For simplicity we consider only  displacements perpendicular to the surface. In the long wave length limit surface acoustic phonons also have a parallel component enforced by the stress free surface condition, but this may be taken into account in the end by a simple numerical factor as discussed in Appendix \ref{sect:A1}. Figure~\ref{fig:Fig1} illustrates the origin of the Dirac electron-phonon coupling. The continuous long wave length deformation of the surface leads to a periodic tilting of the surface normal. The local quasiparticle states have Dirac cones alligned with the local surface normal. We consider only the case of one cone in the surface Brillouin zone. In the continuum limit without inter-cone scattering the effect of several cones would simply be additive. Expanding with respect to the tilt angle one obtains a coupling of the unperturbed quasiparticle states to the surface phonon coordinates.\\ 

In the following we derive the relevant coupling Hamiltonian. Denoting the  displacement vector by \vu({\vvr}) the local direction of the surface normal is given by
\bea
\hn&=&-\frac{u_z'}{(1+u_z'^2)^\frac{1}{2}}\hx + \frac{1}{(1+u_z'^2)^\frac{1}{2}}\hz 
\eea
Approximation to first order in the displacement amplitude leads to
\bea
\hn&\simeq& \hz + \delta\hn \no\\
\delta\hn &=& -u_z'\hx 
\label{eq:normal}
\eea
The first term is the global surface normal $\hz$ and the second the local correction due to the surface displacement where $u_z'=\partial u_z(x)/\partial x$ is the displacement gradient parallel to the phonon wave vector ${\bf q}=q\hq$ where $\hq=\hx$. Expanding in phonon coordinates we get
\bea
u_z'(x)=q\Bigl(\frac{\hbar}{2M\omega_{\vq}}\Bigr)^\frac{1}{2}(b_\vq+b^\da_{-\vq})e^{iqx}
\eea
Inserting $\hn$ into the Hamiltonian of Eq.~(\ref{eq:HL}) we can split into unperturbed $H_0$ and Dirac electron-phonon $H_{ep}$ coupling term. Rearrangement of terms in the latter finally leads to
\bea
&&\\
&&H_{ep}=-\frac{v_F}{N}\sum_{\vk\vq}g_\vq
[(\vk\times\hq)\cdot\hz][\psi^\da_{\vk+\vq}\sigma_z\psi_\vk](b_\vq+b^\da_{-\vq})\no
\label{eq:hep}
\eea
with  the coupling constant defined by 
\bea
g_\vq=q\Bigl(\frac{\hbar}{2M\omega_{\vq}}\Bigr)^\frac{1}{2}\quad\mbox{or}\quad
g_q^2=\frac{1}{2}\Bigl(\frac{\hbar\omega_\vq}{c_sV_c}\Bigr)
\eea
Here $\omega_\vq=v_sq$ is the frequency and $v_s$ the velocity of long wave length
surface phonons. It is related to the elastic constant $c_s=\rho v_s^2$ with $\rho=M/V_c$ denoting the
density and $M, V_c$ the atomic mass and volume per atom respectively. Because
$[(\hk\times\hq)\cdot\hz]=\sin\theta_{\vk\vq}$ where $\theta_{\vk\vq}=\angle (\hat{\vk},\hat{\vq})$
the electron-phonon coupling in Eq.~(\ref{eq:hep}) vanishes for forward scattering ($\theta_{\vk\vq}=0$)
and backward scattering ($\theta_{\vk\vq}=\pm\pi$). In the following we simplify the notation to $\theta_\vk=\theta_{\vk\vq}$ 
because $\hq = \hx$ is fixed.\\

To perform useful calculations the coupling term has to be transformed to the helical Dirac eigenstates of $H_0$ given in Eq.~(\ref{eq:helical}). This is accomplished by applying the unitary transfromation W to $H_{ep}$. We then obtain
\bea
H_{ep}&=&-v_F\frac{1}{N}\sum_{\vk\vq}g_\vq
[(\vk\times\hq)\cdot\hz][\Psi^\da_{\vk+\vq}\hat{\Gamma}_{\vk\vq}\Psi_\vk](b_\vq+b^\da_{-\vq})\no\\
\Psi_{\vk}&=&
\left(
\begin{array}{c}
\gamma_{\vk +}\\
\gamma_{\vk -}
\end{array}
\right);\quad
\hat{\Gamma}_{\vk\vq}=\Gamma_{\vk\vq}\tau_0+\bar{\Gamma}_{\vk\vq}\tau_x
\eea
Here $\tau_0$, $\tau_x$ are unit and Pauli matrix respectively in the space of helical eigenstates. The intra- ($\Gamma$) and inter- ($\bar{\Gamma}$) band transition matrix elements between opposite and equal helicity states respectively are given by ($k_\pm =k_x\pm ik_y$ etc.)
\bea
\Gamma_{\vk\vq}&=&k\sin\theta_\vk\Bigl[\frac{(k+q)_+k_-}{|\vk+\vq|k}-1\Bigr]\no\\
\bar{\Gamma}_{\vk\vq}&=&k\sin\theta_\vk\Bigl[\frac{(k+q)_+k_-}{|\vk+\vq|k}+1\Bigr]
\label{eq:ga}
\eea
The momentum dependent part in the parenthesis may be written as
\bea
\frac{(k+q)_+k_-}{|\vk+\vq|k}&=&
\frac{k+\hk\cdot\vq+i(\hk\times\vq)\cdot\hat{{\bf z}}}
{(k^2+q^2+2\vk \cdot\vq)^\frac{1}{2}}\no\\
&=&\frac{k+qe^{i\theta_\vk}}{(k^2+q^2+2kq\cos\theta_\vk)^\frac{1}{2}}
\eea
In the long wave length case $q/k\ll 1$ we can approximate
\bea
\Gamma_{\vk\vq}=iq\sin^2\theta_k\no\\
\bar{\Gamma}_{\vk\vq}=2k\sin\theta_k
\label{eq:galong}
\eea
This limit corresponds to the physically relevant case for ultrasonic SAW provided that $q\ll k_F=\mu/v_F$ which holds except for the chemical potential very close to the Dirac point.

\section{Surface phonon renormalisation by Dirac particle-hole excitations}
\label{sect:phonon}

In this section we calculate the phonon self energy  and derive the renormalized sound velocity and phonon frequency and in particular the damping due to excitation of surface state conduction electrons. The dependence on wave vector and chemical potential of these quantities is in principle accessible by experiment.

\subsection{Phonon self energies}
\label{subsect:polarisation}

\begin{figure}
\includegraphics[width=80mm]{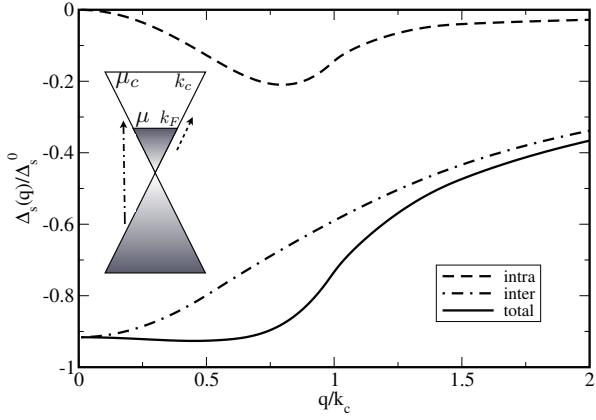}
\caption{Renormalised phonon frequency change $\Delta_s(\vq)=(\tilde{\omega}_{\vq s}-\omega_{\vq s})/\omega_{\vq s}$   for $\mu/\mu_c=k_F/k_c=0.5$. Interband contribution is monotonous while intraband part exhibits a Kohn anomaly for $q/k_c=q/2k_F < 1$ which is also visible in the change of total renormalised phonon frequency. The inset shows a schematic view of intra- and interband excitation processes.}
\label{fig:Fig2}
\end{figure}

The frequency $\tilde{\omega}_{\vq s}$ of surface phonons renormalised by coupling to helical Dirac electrons is obtained from the Dyson equation as
\bea
\tilde{\omega}^2_{\vq s}&=&\omega^2_{\vq s}+2\omega_{\vq s}\Pi_s(\vq,\tilde{\omega}_{\vq s}) \no\\
\Pi_s(\vq,\tilde{\omega}_{\vq s})&=&\Pi(\vq,\tilde{\omega}_{\vq s})+\bar{\Pi}(\vq,\tilde{\omega}_{\vq s})
\label{eq:Dyson}
\eea
where $\Pi_s(\vq,\tilde{\omega}_{\vq s})$ is the total self energy due to surface electrons and $\Pi,\bar{\Pi}$ denote the intra- and inter-band contributions respectively. The effects of the background bulk excitations across the valence-conduction band gap are assumed to be already included in the unrenormalised phonon frequency $\omega_{\vq s}$. In second order perturbation theory with respect to the coupling constant $g_\vq$ the contribution of surface states to the self energy is given by ($\mu > 0$):
\bea
\Pi(\vq,\omega)&=&v_F^2g_\vq^2\sum_\vk |\Gamma_{\vk\vq}|^2
\frac{\theta(\mu-v_F|\vk +\vq |)-\theta(\mu-v_F|\vk |)}
{\omega+v_F(|\vk+\vq |-|\vk |)+i\eta}\no\\
\bar{\Pi}(\vq,\omega)&=&v_F^2g_\vq^2\sum_\vk |\bar{\Gamma}_{\vk\vq}|^2
\frac{-\theta(v_F|\vk +\vq |-\mu)}
{\omega+v_F(|\vk+\vq |+|\vk |)+i\eta}
\label{eq:pol}
\eea
The excitation processes contributing to the self energies are indicated in the inset of  Fig.~\ref{fig:Fig2}. We do not evaluate these
expressions in the general case but  restrict to $\omega \rightarrow 0$ appropriate in the adiabatic limit $v_s\ll v_F$.
Defining  $\Pi(\vq,\omega)=v_F^2g_\vq^2\hat{\Pi}(\vq,\omega)$ etc. we can approximate for phonons with $q/k_F\ll 1$ according to
\bea
\hat{\Pi}(\vq)&=&-\sum_\vk |\Gamma_{\vk\vq}|^2\delta(\mu-v_F|\vk |)\no\\
\hat{\bar{\Pi}}(\vq)&=&-\sum_\vk \frac{|\bar{\Gamma}_{\vk\vq}|^2}{2v_F|\vk |}\theta(v_F|\vk |-\mu)
\eea
Using the long wave length expressions for the matrix elements given in Eq.~(\ref{eq:galong}) we evaluate these expressions as
\bea
\hat{\Pi}(q)&=&-\frac{3}{8}D(\mu)q^2\no\\
\hat{\bar{\Pi}}(q)&=&-\frac{1}{3}\Bigl(\frac{v_F}{\mu}\Bigr)D(\mu)(k_c^3-k_F^3)
\label{eq:pollong}
\eea
where $k_c=\mu_c/v_F$ is a cutoff which corresponds approximately to the wave vector where the surface states merge into the bulk bands. Here $\mu_c$ is one half of the conduction-valence band gap. Furthermore $D(\mu)=\frac{A}{2\pi}\frac{\mu}{v_F^2}$ is the density of states at the chemical potential with $A=a^2$ denoting the surface unit cell area.  Obviously for $q/k_F\ll 1$ the intraband contribution is negligible compared to the interband part because the former involves only few excitation processes around $\mu$. 

%
%

\subsection{Sound velocity correction and attenuation of surface acoustic waves}
\label{subsect:ultrasound}

Then, using Eqs.~(\ref{eq:Dyson},\ref{eq:pollong}) one finally arrives at the surface phonon velocity correction ($\mu > 0$) defined by $\Delta v_s = (\tilde{v}_s - v_s)/v_s$. We obtain
\bea
\Delta v_s&=&
-\frac{1}{6}\bigl(\frac{\mu_c}{c_sV_c}\bigr)\bigl(\frac{D(\mu)\mu_c^2}{\mu}\bigr)
\bigl[1- \bigl(\frac{\mu}{\mu_c}\bigr)^3 \bigr]\no\\
&=&-\frac{\pi}{12}\bigl(\frac{\mu_c}{c_sV_c}\bigr)\bigl(\frac{k_c}{k_0}\bigr)^2
\bigl[1- \bigl(\frac{\mu}{\mu_c}\bigr)^3 \bigr]
\label{eq:velshift}
\eea
where $k_0=\pi/a$ defines the dimension of the surface Brillouin zone.
We can also write $Mv_s^2=c_sV_c$ because $v_s=(\rho/c_s)^\frac{1}{2}$.
Equation~(\ref{eq:velshift}) is a constant velocity shift, independent of  q, but it depends on the Fermi level $\mu$.  It  may therefore be identified as surface contribution by changing the chemical potential $\mu$, e.g. by application of a gate voltage.
In fact it vanishes when the Fermi level reaches the conduction band at $\mu=\mu_c$ because no possible surface interband excitation processes are left in this case.\\

A further telling effect of surface Dirac electrons may be found in the intrinsic attenuation coefficient due absorption of surface phonons by particle-hole excitations. For that purpose the imaginary part of the self energy at small but finite frequencies $\omega=v_sq$ has to be evaluated.
It is obvious from Eq.~(\ref{eq:pol}) and the inset in Fig.~\ref{fig:Fig2} that only intraband processes can contribute to the phonon absorption because $\omega_{\vq s}\ll\mu$. They are given by the imaginary part ($x\equiv q/k$) 
\bea
\hat{\Pi}''(\vq,\omega)&=&\pi\sum_{\vk}|\Gamma_{\vk\vq}|^2\mu x\cos\theta_\vk\no\\
&&\cdot\delta(\mu-v_Fk)\delta(\omega+xv_Fk\cos\theta_\vk)
\eea
For $\omega=v_sq$ this may  be evaluated leading to
\bea
\hat{\Pi}''(q)=A\frac{\mu}{v_F^2}\frac{v_s}{v_F}q^2[1-(\frac{v_s}{v_F})^2q^2]^\frac{3}{2}
\eea
The attenuation rate per length is defined as $\alpha_{\vq s}=\gamma_{\vq s}/\hbar v_s$ where 
$\gamma_{\vq s}=v_F^2g_\vq^2\hat{\Pi}''(\vq)$ and in leading order of q is finally given by ($A=a^2$)
\bea
\alpha_{\vq s}=\frac{1}{2}Ak_F\Bigl(\frac{\hbar\omega}{c_sV_c}\Bigr)q^2=
\frac{1}{2\hbar}A\frac{\mu}{v_F}\Bigl(\frac{\hbar\omega}{c_sV_c}\Bigr)q^2
\eea
This is an intrinsic damping given by phonon absorption due to particle-hole pair creation which does not depend on the mean free path of surface electrons. Its characteristic features are the $q^3$ dependence on wave number (since $\omega =v_sq$) and the linear dependence on the chemical potential or surface density of states. The velocity shift and attenuation of surface phonons in topological insulators can possibly be observed in ultrasonic SAW experiments if the change of $\mu$ around the Dirac point can be achieved by a gate voltage tuning.

%
\begin{figure}
\includegraphics[width=80mm]{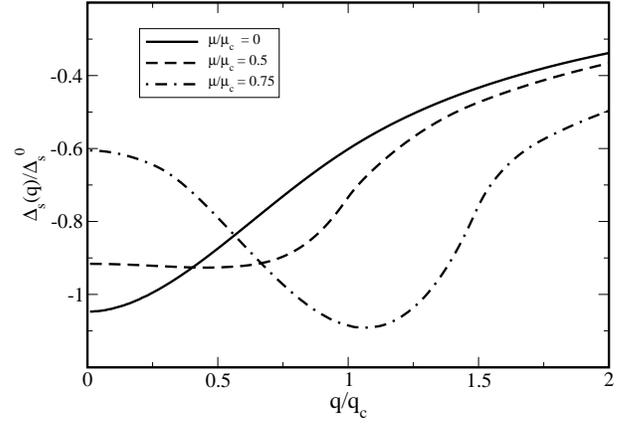}
\caption{The evolution of Kohn anomaly in the phonon frequency change $\Delta_s(\vq)=(\tilde{\omega}_{\vq s}-\omega_{\vq s})/\omega_{\vq s}$  with chemical potential or Fermi wave vector $\mu/\mu_c=k_F/k_c$. For increasing $\mu (k_F)$ the intraband contribution and associated anomaly grows and shifts to larger momentum.
The values for $\vq\rightarrow 0$ show the dependence of sound velocity change $\Delta v_s/\Delta_s^0$  on the chemical potential.}
\label{fig:Fig3}
\end{figure}
%

\subsection{Surface phonon Kohn anomaly for helical Dirac electrons}
\label{sect:kohn}

In this section we consider the renormalisation of surface phonon frequency at finite wave vector comparable to the Fermi wave vector $k_F=\mu/v_F$. Because the self energy will depend on the wave vector one expects a \vq -dependent renormalisation of the phonon frequency which will be noticable in particular around $|\vq|\simeq 2k_F$. In analogy to the case of bulk bands this may be called a Kohn anomaly caused by helical Dirac electrons. Since $k_F\ll k_0=\pi/a$  the background (bare) phonon dispersion $\omega_{\vq s}$ is still linear in $|\vq|$ in this regime. We consider the normalised change in the phonon frequency given by $\Delta_s(\vq)=(\tilde{\omega}_{\vq s}-\omega_{\vq s})/\omega_{\vq s}$. It may be calculated from the static part ($\omega\rightarrow 0$) of the self energies since $v_s/v_F\ll 1$. We assume $\vq=q\hat{\bf x}$ and define dimensionless quantities $\tmu=\mu/\mu_c,\;  \tq=\vq/k_c,\; \tk=\vk/k_c$. After some algebra we then obtain ($\mu >0$)

\bea
\Delta_s(\tq)&=&\Delta_s^0(\phi(\tq)+\bar{\phi}(\tq))\no\\
\phi(\tq)&=&-\int_0^1d\tilde{k}\int_0^{2\pi}d\theta\tilde{k}^3\sin^2\theta\no\\
&&[1-\frac{\tilde{k}+\tilde{q}\cos\theta}{|\tk+\tq|}]
\frac{\theta(|\tk-\tmu|)-\theta(|\tk+\tq|-\tmu)}{|\tk+\tq|-|\tk|}\no\\
\bar{\phi}(\tq)&=&-\int_0^1d\tilde{k}\int_0^{2\pi}d\theta\tilde{k}^3\sin^2\theta\no\\
&&[1+\frac{\tilde{k}+\tilde{q}\cos\theta}{|\tk+\tq|}]
\frac{\theta(|\tk+\tq|-\tmu)}{|\tk+\tq|+|\tk|}
\label{eq:renormvel}
\eea

where $|\tk+\tq|=(\tilde{k}^2+\tilde{q}^2+2\tilde{k}\tilde{q}\cos\theta)^\frac{1}{2}$ and $\theta=\theta_{\vk\vq}$ is the angle between the wave vectors. The momentum integration extends up to the cutoff wave number $k=k_c$ or $\tilde{k}=1$ where the helical surface states merge with the bulk bands. The scale of the q-dependent phonon frequency renormalisation in dimensionless units is given by
\be
\Delta_s^0=\frac{1}{4}\bigl(\frac{\mu_c}{c_sV_c}\bigr)\bigl(\frac{k_c}{k_0}\bigr)^2
\label{eq:scale}
\ee
It also determines the scale of SAW velocity change in Eq.~(\ref{eq:velshift}) with $\Delta v_s^0= (\pi/3)\Delta_s^0$.
We note that the integrands in Eq.~(\ref{eq:renormvel}) vanish for $\theta =\pi$ where the phonon couples to back-scattering processes  $\vk\rightarrow\vk+\vq=-\vk$. This arises because helical states with \vk~ and -\vk~ have opposite spin and the phonon cannot couple to spin-flip processes according to the Hamiltonian in Eq.~(\ref{eq:hep}). In the usual bulk conduction bands where the spins are not coupled to the momentum system it is the  backscattering intraband processes that lead to the 'Kohn anomaly' of renormalised phonons at a wave vector $q=2k_F$. It may be more or less pronounced depending on dimensionality and nesting properties of the bands \cite{Grunerbook}.\\ 

In the present case we have both interband and intraband contributions to the renormalised phonon frequency. The resulting change in the velocity is shown in Fig.~\ref{fig:Fig2} with both contributions shown in addition to the total change for a chemical potential $\hat{\mu}=0.5$  $(k_F=0.5k_c)$, i.e. halfway in the upper Dirac cone. The absolute value of the  interband contribution shows a monotonously decreasing  behaviour as function of $\hq$. On the other hand the magnitude of the intraband contribution has a pronounced maximum. Because of the suppression of backscattering it is, however, at an intermediate wave vector $q < 2k_F = k_c$  between forward (q=0) and backscattering  ($q=2k_F$). As a result the total renormalised velocity change shows a resulting Kohn anomaly appearing at  $q<2k_F$.\\ 

This anomaly in the surface phonon frequency change depends on the chemical potential or Fermi wave vector. It is shown in Fig.~\ref{fig:Fig2} for three different cases. When $\mu=0$ there is no intra-band contribution and one has monotonous q-dependence. For increasing $\mu$ the Kohn anomaly progressively appears and shifts to higher q-values. However the maximum effect is always seen at $q < 2k_F$ due to the lack of backscattering. The absolute magnitude of the anomaly is determined by Eq.~(\ref{eq:scale}). Using  parameter estimates for \BS~  according to $\mu_c\simeq 0.3$ eV, $c_sV_c\simeq 4.68$ eV and $k_c/k_0\simeq 0.1$ we get the dimensionless anomaly scale $\Delta_s^0\simeq 1.6\cdot 10^{-4}$. It is largely determined by the small value of $(k_c/k_0)^2\simeq 0.01$ which is the small relative area of the Dirac cone at $\mu=\mu_c$ compared to the surface Brillouin zone \cite{Zhang10}.
Although $\Delta_s^0$ is not very large it is perfectly sufficient for the observation of elastic constant anomalies (as function of $\mu$) since for the latter relative changes can be measured with a  precision up to $10^{-6}$. On the other hand observing a Kohn anomaly at this scale is a challenge. In the continuum Dirac model there is  no nesting involved in the 2D Fermi surface. Therefore there is no singular enhancement of the Kohn anomaly. However if one includes warping effects \cite{Zhou09,Fu09} due to higher order terms in the surface cone dispersion which are important at larger chemical potential nesting effects appear between various symmetry {\vk}- points (not connected by backscattering). This could lead to a considerable enhancement of the Kohn anomaly at these nesting vectors.

\section{Conclusion}
\label{sect:conclusion}
 In this work we investigated the effect of helical Dirac electrons in topological insulators on the propagation of surface phonons. Their coupling to the quasiparticle-hole excitations has been derived within a continuum model assuming a displacement dependent dispersion. The coupling to  Dirac cone quasiparticles renormalizes the sound velocity and leads to intrinsic damping of surface acoustic waves. The size of these renormalisation effects depend directly on the chemical potential and may therefore be checked  e.g. by application of a gate voltage or opening a gap through magnetic doping \cite{Chen10a} in surface ultrasonic experiments. Furthermore at finite wave vectors the surface phonon dispersion exhibits Kohn anomalies. Because of the vanishing backscattering due to helical spin polarization they appear at wave vectors $q <2k_F$ and their magnitude depends on the position of the chemical potential. Due to the absence of nesting effect in the continuum limit they show no singular q-dependence. The latter may change if warping effects are included for higher chemical potential. It will be an experimental challenge to identify these predicted anomalies which may also be tuned by application of gate voltage.\\

\section*{Acknowledgements}

The author would like to thank Lei Hao for helpful discussion and comments.

\appendix
\section{}
\label{sect:A1}

Our discussions of surface acoustic phonon coupling were simplified by ignoring the longitudinal displacement field that exists for surface waves due to the stress free boundary conditions in the limit ${\bf q}\rightarrow 0$. We briefly discuss here the modifications caused  by the inclusion of the longitudinal displacement part $u_x(x)$. In its presence the modification to the surface normal in Eq.~(\ref{eq:normal}) is changed to
\bea
\delta\hn &=& -u_z'[1+2\frac{u'_x}{u'_z}\frac{u''_x}{u''_z}]\hx =-u'_z[1-2\gamma_s]\hx
\label{eq:normal2}
\eea
where prime and double prime denote first and second derivatives with respect to x (propagation direction) respectively. Evaluation of the derivatives 
for a Rayleigh-type long wave length surface phonon leads to
\bea
\gamma_s=\frac{\xi^2(1-\xi^2)^\frac{1}{2}}
{(2-\xi^2)-2(1-\xi^2)^\frac{1}{2}(1-\frac{v_t^2}{v_l^2}\xi^2)^\frac{1}{2}}
\eea
Here $v_{l,t}$ are longitudinal and transverse bulk sound velocities and $\xi=v_s/v_t$ where $v_s$ is the surface phonon sound velocity. Thus inclusion of the longitudinal displacement field merely modifies the electron phonon coupling constant in Eq.~(\ref{eq:hep}) by a factor $(1-2\gamma_s)$.

\end{document}